\documentclass[a4paper,11pt]{article}
\usepackage{pos}

\usepackage{subcaption} 
\usepackage[LGR,T1]{fontenc}
\usepackage[greek, english]{babel}
\usepackage{alphabeta}

\babelprovide[import=el, onchar=ids fonts]{greek}

\usepackage{lineno}

\title{Determination of the sensitivity of the DEAP-3600 experiment to supermassive charged gravitinos}

\author{Michał Olszewski for the DEAP Collaboration}

\affiliation{AstroCeNT, Nicolaus Copernicus Astronomical Center of the Polish Academy of Sciences,\\
  ul. Rektorska 4, Warsaw, Poland}

\emailAdd{molszewski@camk.edu.pl}

\abstract{The lack of discovery of particle dark matter candidates within the favored mass-window range brings in the motivation for the study of new options brought by Planck-mass dark matter models. Extended supergravity theories predict the existence of non-relativistic gravitinos that could at least in part contribute to the missing mass-energy density of the Universe. The feasibility study for the discovery with DEAP-3600 experiment of  Planck mass charged gravitino dark matter is presented.
Additionally the expected signal events topology within the detector is discussed.}

\FullConference{19th International Conference on Topics in Astroparticle and Underground Physics (TAUP2025)\\
24–30 Aug 2025\\
Xichang, China\\}

\begin{document}
\maketitle

\section{Introduction}

The existence of dark matter (DM), which constitutes about 27\% of the Universe’s mass-energy content, is strongly supported by multiple lines of observational evidence, including galaxy rotation curves and cosmic microwave background measurements~\cite{Zwicky1933, Rubin1980, Planck2020}. It is characterized as non-baryonic, non-luminous, cold, and collisionless, yet its precise nature remains unknown since no particles within the Standard Model (SM) satisfy all necessary properties~\cite{Copi1996, Burles1998}. Among the favored particle candidates are Weakly Interacting Massive Particles (WIMPs), such as the neutralino predicted by supersymmetric theories, as well as axions and other particles from other SM extensions~\cite{Jungman1996, ArkaniHamed2002}. 
Alternatives to particle candidates also exist, with primarily examples being primordial black holes and modified gravity theories~\cite{Carr2016, Milgrom1983}.
In this paper, however, we would like to follow recent work~\cite{Kruk_2025} proposing DM consisting at least in part of an extremely dilute gas of supermassive stable gravitinos. 
This idea originates from Gell-Mann’s insight that the SM three generations of fermions correspond to the spin-$\frac{1}{2}$ states of the maximal N=8 supermultiplet once eight Goldstinos are removed. The only additional fermions predicted are eight massive gravitinos.
It is proposed~\cite{meissner2023stablesupermassivechargedgravitino}, that under $SU(3)\times U(1)_{em}$ symmetry group they would split as
\begin{equation*}
    \left( \mathbf{3}, \frac{1}{3} \right) \oplus \left( \mathbf{\bar{3}}, -\frac{1}{3} \right)
    \left( \mathbf{1}, \frac{2}{3} \right) \oplus \left( \mathbf{1}, -\frac{2}{3} \right),
\end{equation*}
therefore, making such DM candidates carry a fractional electric charge. Apart from theoretical considerations, such a mechanism ensures stability of the model as superheavy particles participating in standard (electromagnetic) interactions would otherwise decay at a very early stage in the evolution of the Universe~\cite{PhysRevD.100.035001}.

Extended supergravity theories, such as the one considered here, are interesting in their own right.
Although their experimental verifiability is uncertain, they are nevertheless well motivated theoretically. On the one hand, they incorporate infinite-dimensional duality symmetries into unification in novel ways, and on the other, they justify the lack of evidence of new physics beyond the electroweak scale. The indication that the SM might survive up to the Planck scale is an incentive for the study of new options brought by  Planck-mass charged gravitino dark matter models~\cite{PhysRevD.100.035001}.

The lack of discovery of gravitino DM may be explained by the implied features of these particles. Apart from being extremely massive with masses $m \approx M_{Pl}$, their predicted abundance in the Universe is very low, of the order of $3 \times 10^{-14}$ particles per cubic meter. As a consequence of this, and of the small cross-section for interaction, gravitino-antigravitino annihilations, although possible, are extremely rare. Similarly, interaction with CMB may be neglected due to suppression of the cross-section by the Planck length scale factor. Furthermore, DM gravitinos are never in thermal equilibrium during the evolution of the Universe after the Planck era, hence their velocity distribution is unknown. Typically, it is assumed that the average velocity of superheavy DM particles w.r.t. the Earth is between $v_E \sim 30$ km/s and $v_S \sim 230$ km/s, which are the virial velocities for a particle bound to the Sun and to our Galaxy, respectively. Due to the fractional charge, DM gravitinos are stable and interact with SM matter by uniformly exciting or ionizing their surroundings, leaving a straight track all along their path. The estimated flux is approximately equal to 0.003 m$^{-2}$ yr$^{-1}$ sr$^{-1}$~\cite{PhysRevD.100.035001}.

\section{DEAP-3600 experiment}

The properties of SM gravitinos, together with their assumed detector signature, determine the most suitable experimental setup for detection. The accelerator experiments are characterized by their very high background and typically are based on periodic triggers (e.g., 25~ns for LHC detectors), which in practice prevents measurements of a weak DM signal not falling into the trigger synchronization window. Similar restrictions apply to large neutrino experiments (e.g., Super-Kamiokande) where the trigger is operating in the regime of relativistic particles. Additionally, those experiments are limited to Cherenkov radiation detection, which does not occur for non-relativistic particles like gravitinos. 
On the other hand, past detectors like MACRO (originally designed for magnetic monopole detection) or DM liquid xenon (LXe) detectors (LUX, XENON1T), are too small or optimized for different signals, resulting in very low expected event rates for gravitinos.
The current biggest LAr and LXe experiments - DEAP-3600, LZ, XENONnT and Panda-4T - present similar theoretical sensitivity with respect to gravitinos\footnote{As the size of the detector is crucial for detection of gravitinos the future LAr-based experiments will have an advantage over LXe-based ones.}.
The most promising experiments for detecting supermassive charged gravitino DM are large underground neutrino detectors such as JUNO and the upcoming liquid argon-based (LAr) DarkSide-20k and DUNE.

Knowing the estimated flux of Planck mass DM gravitinos and the active volume of a given detector, we are able to calculate the number of signal events over a given time of exposure. 
Ref ~\cite{PhysRevD.100.035001} predicts 0.009 and 0.05~events for LUX and XENONnT experiments for a total available data exposure time of 427 and 316~days, respectively. 
For upcoming large LAr experiments, we get 0.2 and 1.6~events per year for the fiducial volume of DarkSide-20k and ARGO detectors, respectively. For DEAP-3600 the expected signal rate is equal to 0.15 events per 820 days of Run II dataset.

The DEAP-3600 experiment, located 2~km underground at SNOLAB in Sudbury, Canada, uses 3.3~tonnes of ultra-pure liquid argon (LAr) as a target for DM detection. Its central component is a spherical acrylic vessel with an 85~cm radius, coated on the inside with a 3~$\mu$m layer of tetraphenyl butadiene (TPB), which shifts the ultraviolet argon scintillation light (128 nm) to visible light ($\sim$420~nm) detected by 255 high-efficiency Hamamatsu photomultiplier tubes (PMTs). The detector is surrounded by a stainless steel shell immersed in an ultra-pure water tank acting as a Cherenkov muon veto, significantly reducing cosmic muon background. DEAP-3600’s sophisticated data acquisition system digitizes the PMT signals with high precision to discriminate scintillation events. The experiment has operated continuously since 2016. Its large target mass, low-background environment, and efficient shielding make DEAP-3600 one of the most sensitive detectors for WIMP DM searches, achieving leading exclusion limits in LAr and providing a platform for R\&D for larger future detectors.

Pulse-shape discrimination (PSD) method utilized in DEAP-3600 searches leverages the distinct scintillation timing properties of argon to differentiate nuclear recoils (NR), expected from DM interactions, from electronic recoils (ER), mainly background beta and gamma events. This technique exploits the different lifetimes of argon excimers in singlet ($\sim 6$ ns) and triplet ($\sim 1.5 \mu$s) states, which are predominantly excited by NR and ER, respectively, leading to characteristic pulse shapes. The PSD parameter, so-called Fprompt, often used is the ratio of prompt light (within 60~ns) to total light collected, allowing ER suppression by factors exceeding 10$^7$ to 10$^{10}$ above detection thresholds. DEAP-3600’s high-efficiency photomultiplier tubes and precise waveform digitization enable effective pulse shape analysis with low leakage of ER into the NR acceptance region.

\begin{figure}[htbp]
  \centering

  \begin{subfigure}{0.43\linewidth}
    \centering
    \includegraphics[width=\linewidth]{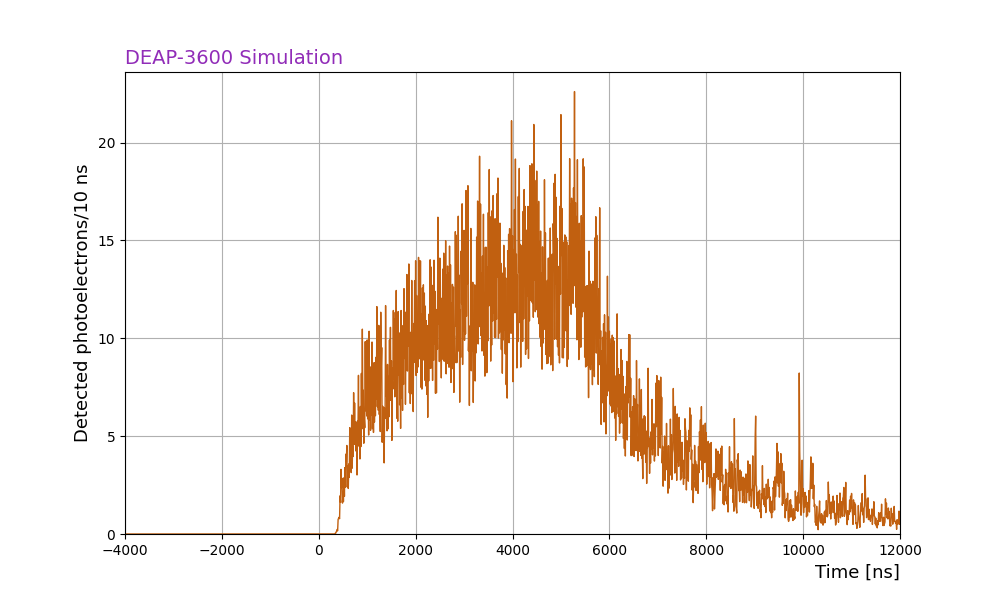}
    \label{fig:bottom_left}
  \end{subfigure}
  \hfill
  \begin{subfigure}{0.53\linewidth}
    \centering
    \includegraphics[width=\linewidth]{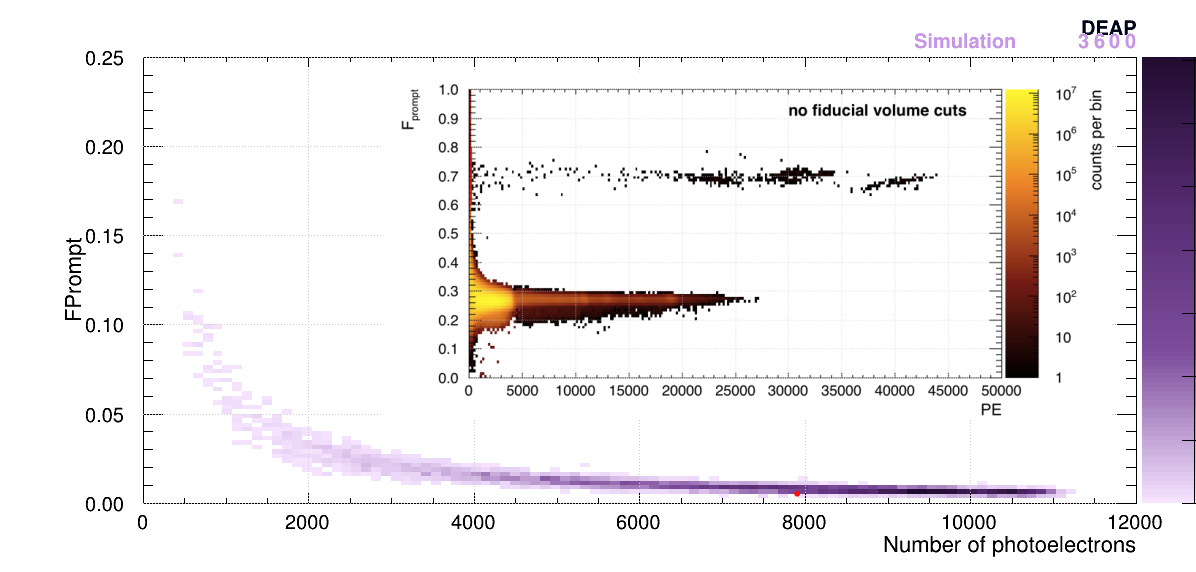}
    \label{fig:bottom_right}
  \end{subfigure}

  \caption{(left) Sample reconstructed waveform for a simulated event with incoming superheavy gravitino millicharged DM particle passing 1169 mm of fiducial volume with calculated $F_{prompt} = 6\times 10^{-3}$; (right) Distribution of Fprompt vs. number of photoelectrons in the simulated events with embedded distribution for unfiltered 4.4 live days DEAP-3600 dataset. Red dot indicates position of the event used to produce the waveform plot.}
  \label{fig:three_images}
\end{figure}

Energy reconstruction in the DEAP-3600 experiment relies on the detection of scintillation light produced by particle interactions in the liquid argon (LAr) target. The energy deposited by an event is quantified by counting the total number of photoelectrons (PEs) in the PMTs. The conversion from the observed PE count to recoil energy accounts for the light yield of the detector, which for DEAP-3600 is approximately 7.14~PE
\footnote{Technically speaking, here we define PE as total charge in the waveform after gain correction divided by a charge weighted mean of the average charge for a single PE for the PMTs contributing to the sum.}
per keV electron equivalent (keV$_{ee}$) \cite{Adhikari_2021}. Detailed calibration procedures enable accurate modeling of the detector response. Corrections for position-dependent variations in light collection are applied using advanced position reconstruction algorithms that analyze the spatial and temporal distribution of detected photons.  Collectively, these methods enable DEAP-3600 to reconstruct recoil energies with sufficient accuracy and resolution to effectively discriminate between background and potential DM signals, supporting robust DM searches with low energy thresholds and well-understood systematics.

The background rejection power of Fprompt and reconstructed energy justify their usage as two-dimensional signal variables in the vast majority of physics searches in this experiment. The gravitino DM analysis is no exception here, and from the results presented in~\cite{Kruk_2025} we can already draw some conclusions regarding the appearance of the events for this type of signal. Firstly, it is shown that in the case of a gravitino passing through liquid argon, the nuclear recoil can be neglected 
and the track is build as a straight line comprised of argon excitation events. This translates into the expectation that Fprompt for such signal events would lie in the ER band of the plot. Secondly, since calculated mean times between the electronic recoil excitation events in LAr are 228~ns and 0.12~ns for $v_E$ and $v_S$ respectively~\cite{Kruk_2025}, we calculate the number of produced scintillation electrons to be up to 250 and 62000, depending on the gravitino velocity and presumed track length. This means that with the light yield of the detector the number of PE would be in a range between $10^2$ and $10^{4}$. Both numbers are well inside DEAP-3600's PMTs acceptance window, however, the former one is very small and considering the wide signal distribution within the trigger window (see left plot on Fig.~\ref{fig:three_images}) brings us to the first important conclusion that reconstruction of gravitinos for velocities about $v_E$ is not possible in DEAP-3600 as the signal would be lost in the background or be excluded by the trigger.

\section{Simulation results}

To simulate signal events for the DEAP-3600 experiment, we created superheavy charged gravitino generator using the DEAP-3600's Geant4-based Monte Carlo and analysis tool. We simulated LAr scintillation with 128~nm photons, generated in the detector's fiducial volume in random directions. 
Photons were isotropically emitted from interaction points randomly distributed along the track length, based on the calculated mean free paths.
On top of that, the detector temporal response was emulated using the DM velocity and LAr scintillation time constants. 

As mentioned in the previous section, a gravitino moving with velocity $v=v_E$ cannot be detected with DEAP-3600, hence we will consider only the scenario where $v=v_S$. Firstly, we generated several events to find out what waveforms induced by passing gravitinos would look like.
The outcome, with the example shown on the left plot of Fig.~\ref{fig:three_images}, can be understood in terms of reasoning presented already in the referential Planck mass Multi-scattering Heavy Dark Matter (MIMP) analysis at DEAP-3600 from 2021~\cite{Lai_2021}. Similarly to that previous case, signal is distributed in about 6 $\mu$s window which is related to the time needed to pass through the inner vessel of the detector for the particle with an average velocity $v=300$ km/s. Also the scintillation light in the event is widely uniformly distributed through all the PMTs looking at the inner vessel producing uniformly distributed pulses along the waveform and triggered as a single event. The main difference with respect to the MIMP analysis stems for the origin of the scintillation light. While here it is produced by ER events, in the MIMP analysis the interactions were dominated by NR interactions. The consequence of the slower ER LAr response function is comparatively milder slope of the pulse-shape (see Fig. \ref{fig:three_images}), which in turn translates into even less light in prompt window and pushes the Fprompt to lower values compared to MIMP paper.
Furthermore, we generated 10000 events, the resulting signal plot distribution is shown on the right plot in Fig.~\ref{fig:three_images}. The obtained Fprompt values are in a range between 0 and 0.1. 
These values are in the region of relatively low event rate on the signal plot (see embedded picture on right plot on Fig. ~\ref{fig:three_images}) for dataset with no fiducial volume cuts. The main sources of the event rate coming from LAr bulk $^{39}$Ar $\beta$ decays and the $^{232}$Th and $^{235}$U  chains in the PMTs typically give signal with higher Fprompt values.

\section{Summary}

The sensitivity to supermassive charged gravitinos for the DEAP-3600 experiment has been qualitatively presented. Although the dataset present currently allows only to set exclusion limits on the high-velocity gravitino cross section, due to unique signatures, detection would be within reach of the next generation detectors, e.g., DarkSide-20k and ARGO. Due to large dimensions, design of the trigger system, and utilization of the PSD method,  LAr detectors present a significant sensitivity advantage. 
A gravitino search analysis on the DEAP-3600 will be performed using the simulated signal model presented in this work.

\section{Acknowledgements}
This work was directly supported by the Polish National Science Centre grant ~(UMO-2024/08/X/ST2/00430).
We thank the Natural Sciences and Engineering Research Council of Canada (NSERC),
the Canada Foundation for Innovation (CFI),
the Ontario Ministry of Research and Innovation (MRI), 
and Alberta Advanced Education and Technology (ASRIP),
the University of Alberta,
Carleton University, 
Queen's University,
the Canada First Research Excellence Fund through the Arthur B.~McDonald Canadian Astroparticle Physics Research Institute,
SECIHTI Project No. CBF-2025-I-1589,
DGAPA UNAM Grants No. PAPIIT IN105923 and IN102326,
the European Research Council Project (ERC StG 279980),
the UK Science and Technology Facilities Council (STFC) (ST/K002570/1 and ST/R002908/1),
the Leverhulme Trust (ECF-20130496),
the Russian Science Foundation (Grant No. 21-72-10065),
the Spanish Ministry of Science and Innovation (PID2019-109374GB-I00) and the Community of Madrid (2018-T2/ TIC-10494), 
the International Research Agenda Programme AstroCeNT (MAB/2018/7) funded by the Foundation for Polish Science (FNP) from the European Regional Development Fund,
and 
Studentship support from
the Rutherford Appleton Laboratory Particle Physics Division,
STFC and SEPNet PhD is acknowledged.
We thank SNOLAB and its staff for support through underground space, logistical, and technical services.
SNOLAB operations are supported by the CFI
and Province of Ontario MRI,
with underground access provided by Vale at the Creighton mine site.
We thank Vale for their continuing support, including the work of shipping the acrylic vessel underground.
We gratefully acknowledge the support of the Digital Research Alliance of Canada,
Calcul Qu\'ebec,
the Centre for Advanced Computing at Queen's University,
and the Computational Centre for Particle and Astrophysics (C2PAP) at the Leibniz Supercomputer Centre (LRZ)
for providing the computing resources required to undertake this work.

\end{document}